\documentclass[11pt]{article}
\usepackage{hyperref}
\pdfoutput=1

\begin{document}
\title{Wake interaction of two disks falling in tandem}
\author{N. Brosse, S. Cazin and P. Ern \\
\vspace{6pt} Institut de M\'ecanique des Fluides de Toulouse, France}
\maketitle


\begin{abstract}
The fluid dynamics video illustrates the interaction of two disks falling in tandem at Reynolds number close to 100. Two fluorescent dyes were used to visualize the wake of each body. We can observe that the trailing body accelerates thanks to the entrainment provided by the wake of the leading body and eventually catches up the leading body. Then, thick disks (d/t = 3) lose their initial wakes, separate laterally and fall side by side. On the contrary, the wakes of thinner disks (d/t = 10) merge in a single wake and the bodies continue their fall together adopting a stable Y-configuration.
\end{abstract}


We investigated experimentally the interaction of two identical bodies falling in tandem in a fluid at rest. The disks have various diameter-to-thickness ratios (d/t = 3 and 10) and a density close to the fluid one  \cite{Fernandes_2007}.  We focus here on the case of intermediate Reynolds numbers (close to 100) for which the bodies follow a  rectilinear path when they fall alone and a steady axisymmetric wake  develops behind the body. The motion of the bodies was recorded by a travelling camera and the wakes of the disks  were visualized with two different dyes (rhodamine and fluorescein) illuminated with ultraviolet light. The fluid  dynamics videos (high quality and low quality) are taken in the reference frame of the leading body (the first body released). We can observe that the trailing body accelerates thanks to the entrainment provided by the wake of the leading body and eventually catches up the leading body. Then, thick disks (d/t = 3) lose their initial wakes, separate laterally and fall side by side. On the contrary, the wakes of thinner disks (d/t = 10) merge in a single wake that stays attached to the bodies. They thus continue their fall together adopting a stable Y-configuration. This configuration was already observed by \cite{Joseph_1993} and \cite{Joseph_1987} and called a stable wake architecture, since the bodies appear to be glued by the wake. In parallel to the flow visualization, the motion of the bodies was measured thanks to two perpendicular travelling cameras (three-dimensional trajectography). In the case of thick bodies, the final side-by-side configuration is repulsive and the bodies continue to separate while falling until they reach a horizontal separation distance of about 6 diameters. In more than $90\%$ of the cases, the Y-configuration adopted by the thin bodies is stable. This configuration allows the bodies to fall more rapidly ($1.15$ times their velocity in the isolated case). Moreover, the Y-configuration falls along a path which is slightly tilted relative to the vertical (in between $4\%$ and $10\%$ of horizontal drift). At higher Reynolds numbers, the Y-configuration displays a regular transversal oscillation in addition to the horizontal drift and a periodic shedding of vortices in phase with the oscillatory motion is observed. It appears that the two bodies behave like a single rigid body. Even at higher Reynolds numbers for which the bodies display a periodic zigzag path when they fall alone, the entrainment process leads to a Y-configuration. However, at variance with the previous case, the relative distance and relative inclination of the bodies fluctuate in time, so that the bodies may adopt temporarily a T-configuration and do no more behave like a single rigid body.



\begin{thebibliography}{10}

\bibitem[1]{Fernandes_2007}
Fernandes, P., Risso, F., Ern, P. and Magnaudet, J., 2007. Oscillatory motion
and wake instability of freely rising axisymmetric bodies. J. Fluid Mech. 573, 479-502.

\bibitem[2]{Joseph_1993}
Joseph, D., 1993. Chapter 10: Finite size effects in fluidized suspension experiments. 
Particulate Two-Phase Flow, Edited by M. C. Roco. Butterworth-Heinemann.

\bibitem[3]{Joseph_1987}
Joseph, D., Fortes, A. and Singh, P., 1987. Nonlinear Mechanics of Fluidization of Spheres, Cylinders and Disks in 
Water. Advances in Multiphase Flow and Related Problems, Edited by G. Papanicolau, SIAM.

\end{thebibliography}
\end{document}